\documentclass[aps,twocolumn,showpacs,preprintnumbers,amsmath,amssymb]{revtex4}
\usepackage{graphicx}
\usepackage{dcolumn}
\usepackage{bm}

\begin{document}
\def\ltsimeq{\,\raise 0.3 ex\hbox{$ < $}\kern -0.75 em
 \lower 0.7 ex\hbox{$\sim$}\,}
 \def\gtsimeq{\,\raise 0.3 ex\hbox{$ > $}\kern -0.75 em
  \lower 0.7 ex\hbox{$\sim$}\,}

\def\avg #1{\langle #1\rangle}
\title{Looking at map networks with time delay interactions from a local perspective}

\author{A. Parravano$^1$}

\affiliation{$^1$Centro de F\'{\i}sica Fundamental,
Facultad de Ciencias, Universidad de Los Andes, \\ Apartado Postal 26
La~Hechicera, M\'erida~5251, Venezuela,}

\date{\today}

\begin{abstract}
The evolution of  
networks of coupled chaotic maps with delayed interactions can be studied
in the usual way by analyzing the evolution of the state of elements at  
each iteration time (the ``Simulator" point of view),
or it can be analyzed from the point of view of a single element 
(the ``Observer" perspective) that is
receiving delayed information from the other elements in the system.
In the usual``Simulator" timeframe, an absolute time (i.e. the number of iterations $t$)
is adopted to define the system state at each time $t$.
In the``Observer" framework, the state of an element in the system is given by its 
state at the ``Simulator" time $t-\tau$, where $\tau$ is the information travel time
between that element and the Observer.
We emphasize the convenience of the analysis of the system dynamics in the
``Observer" timeframe 
by showing that the ``Observer" dynamics differs substantially from the 
one in the ``Simulator" timeframe, and that the system dynamics 
in the ``Observer" timeframe reflects the proper causality of the interactions
among the elements.
\end{abstract}
\maketitle

Delayed interactions are present in most physical,
biological, and social systems. 
Networks of interacting elements have been extensively used as a first approach 
to modeling these systems. Most of these models have excluded time delays. 
Since the pioneering studies on coupled map lattices by Kaneko and
by Waller and Kapral \cite{{Kan84},{WallerKapral}}, 
such systems have been investigated for a variety
of local dynamics and topologies (from coupling with nearest neighbors to
global coupling \cite{Kan1}, from random to small-world \cite{Watts98} to scale-free 
\cite{Barabasi99} connections.
Many investigations have been devoted to study
the various kind of organized behavior that occurs in a variety of networks of
chaotic elements. 
Synchronization, pattern formation, dynamical clustering, 
and a zoo of organized spatiotemporal
patterns have been observed. Most of these investigations have neglected
the delay due to finite transmission and processing speed among the interacting
elements.   
The inclusion of time delays in these systems increases their dimensionality
and as consequence, a new variety of complex behaviors are accessible.
Although, in real systems several kinds 
of interactions are often acting simultaneously with different propagation and processing
velocities, one of them usually dominates. Then, 
in a first approximation,  many of these systems can be modeled 
as a system of elements interacting with a single velocity 
of transmission. The topology of the network determines the 
transmission path distance, and therefore the time delay between elements
and the possible attenuation of the signal.

The use of maps for the local dynamics of the system elements has
shown that non-local interactions with time delays
can induce a variety of organized behaviors
\cite{{Chapeau92},{Jiang00},{Masoller03},{Fatihcan04}} .
However, the organized behavior can be missed or misunderstood 
if the system is not observed in an adequate way. 
For example, if it would be possible to construct a picture
of the universe in which each point shows its state at an absolute time measured
from the Big Bang, this picture would look very different than the view of a 
real observer who is constrained to see the delayed picture 
(the ``Observer" timeframe).
In this paper we intend to show through a particular example the convenience of 
looking at extended dynamical systems with delayed interaction from the 
``Observer" timeframe.

Another aspect that we intend to demonstrate is the convenience of focusing on the
behavior of the coupling field.
In an autonomous system the coupling field on an element is a function of the system
history and the topology of the connections among the system elements.
It has been recently shown \cite{{PC},{PC2},{CP2000}} that
there is no difference between the 
evolution of an element in an autonomous system and the evolution of an 
isolated element subject to an external forcing that mimics the coupling field.
Thus, the dynamics of the
coupling field on a given element contains the relevant information about the
organization of the system.

A coupled map network (CMN) can be defined as
\begin{equation}
x^i(t+1)=(1 - \epsilon^i) f(x^i(t)) + \epsilon^i \Phi^i(t)  , \,\,    (1\ge i \ge N) \,,
\label{cml}
\end{equation}
where
$\Phi^i(t)$ is the local field at time $t$ and place $i$ that acts as 
a forcing on element $i$ with an efficiency $\epsilon^i$, and $f$ is the local
dynamics.
The functional form of $\Phi$ depends on the topology and on the properties
of the interactions between the elements of the system.
We focus here on $\Phi$ functions of the form
\begin{equation}
\Phi^i(t)= \sum _{j \in \nu_i} \eta _{ji} \, x^j(t-\tau _{ji})  \,,
\label{phi}
\end{equation}
where $\nu_i$ is the set of neighbors of $i$, $\eta _{ji}$ represents the 
attenuation or amplification of the signal in its travel
from place $j$ to place $i$, and $\tau _{ji}$ is the delay.
As an example, here we consider $\Phi$ of the form
\begin{equation}
\Phi^i(t)=\frac{
\sum _{l=1}^{N_v} l^{-\alpha} [ x^{i+l}(t-m l) + x^{i-l}(t-m l)]}
{2 \sum _{l=1}^{N_v} l^{-\alpha} }  \,,
\label{phi_ret}
\end{equation}
where, for a one dimensional array of equally spaced elements, $N_v$ is the number 
of neighbors at each side of element $i$ that are coupled to it, $l=1,..,N_v$ is the 
distance to $i$, 
and $m l$ is the time delay (i.e. $m$, assumed to be an integer $\ge 0$, is the inverse 
of the signal velocity, or the information time processing per site). 
The parameter $\alpha$ is an attenuation parameter
(i.e. $\alpha=0,1,2$ represents the flux decrease of a signal in a one, two
or three dimensional space; $\alpha$ negative implies amplification).

This CMN model, with periodic boundary conditions, is used in what follows 
to show the evolution of the
system when it is seen from the ``Simulator" and from the ``Observer" timeframes.
Hereafter, by Simulator timeframe we mean that
an absolute time (i.e. the number of iterations $t$)
is adopted to define the state of the system.
In the Observer timeframe, the state of a given element $j$ in the system,
from the point of view of the observer at $i_{obs}$ is given by the state of element $j$
at ``Simulator" time $t-\tau _{ji_{obs}}$, where $\tau_{ji_{obs}}$ is the
information travel time between that element and the observer.
Suppose that an array of elements on a line display a ``chess-board-like''
spatiotemporal pattern in the Simulator timeframe. Then, for $m$ odd,
this system will display in the Observer timeframe a ``strips-like'' 
spatiotemporal pattern.
For the "Simulator", the mean value $<x>_{\rm Sim}(t) \propto \sum x^i(t)$ 
and the spatial dispersion 
$\sigma_{\rm Sim}(t) \propto \sum [x^i(t)-<x>_{\rm Sim}(t)]^2$ are 
stationary (if $N$ is even).
In contrast, for the "Observer", the mean value 
$<x>_{\rm Obs}(t) \propto \sum x^i(t-m|i_{obs}-i|)$ is
periodic and the spatial dispersion
$\sigma_{\rm Obs}(t) \propto \sum [x^i(t-m|i_{obs}-i|)-<x>_{\rm Obs}(t)]^2$ 
is zero.

Patterns in the evolution of dynamical systems can be detected by
direct observation of the system spatiotemporal states, from their power 
spectra and from quantitative measurements of the mutual correlation among
system elements \cite{{PC},{GA90},{AG91}}. These techniques of analysis are
obviously consistent among each other but they differ on the their ability 
to detect and quantify different aspects of the patterns. 
However, any of these methods of analysis
give in general different results depending on the representation timeframe.
Spatiotemporal organized patterns can also manifest
in the time dependence of some order parameter such as the mean value, 
dispersion, number of domains, length of interphase, etc. But again, 
the evolution of these order parameters depend in general on the used 
timeframe representation.
It is a direct task to define the Observer timeframe if the system dynamics
(i.e. Eqs. (1-3)) is known, but without having some clues on the subjacent 
dynamics the derivation of the appropriate Observer timeframe 
from the unique analysis of the spatiotemporal pattern in the Simulator 
timeframe is in general complicated if not ambiguous or impossible.
It is out of the scope of this short paper to develop such a method, but
as mentioned above, an important clue on the subjacent
dynamics can be obtained if the coupling field $\Phi^i(t)$ can be measured.
The main point we want to stress here is that different interpretations
of the organized behavior of a system are obtained depending on adopted
timeframe, and that the Observer timeframe provides the simplest representation
to quantify the interaction of a system element with its coupling field.

We show results for
a chain of $N=60$ elements with periodic boundary conditions, 
whose local dynamics is given by the 
logistic 
map $f(x)=1-rx^2$ with $r=2$.
The coupling is uniform (i.e. $\epsilon^i=\epsilon$),
the delay parameter is $m=1$, the signal intensity
decays as $1/d$ (i.e. $\alpha=1$),
and the number of coupled neighbors is $N_v=30$.

Figures 1(a) and 1(b) show, respectively, the
dependence of the spatial dispersion with the coupling strength in the 
Simulator and in the Observer timeframes, for the same random initial conditions
for all $\epsilon$. For $\epsilon\ge 0.55$ the spatiotemporal patterns are flat
(constant in space and time).
For $0.5 > \epsilon > 0.55$ the patters in the
Simulator timeframe is ``strips-like'' and in the Observer timeframe is
``chess-board-like''; 
note that the ``strips-like'' pattern in the Simulator timeframe
is interpreted as syncronization.
Between $\epsilon = 0.13$ and $0.18$ the reverse occurs: the pattern are
quasi ``strips-like'' in the Observer timeframe and quasi ``chess-board-like''
in the Simulator timeframe. 
Below $\epsilon = 0.12$ the patters are turbulent in both timeframes.
\begin{figure}
 \includegraphics[width=8cm]{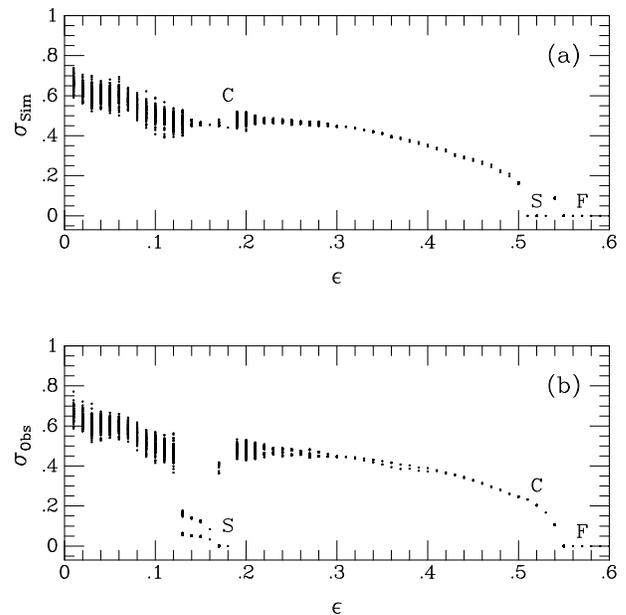}
\caption{ Dependence of the spatial dispersion with the coupling strength in
a) the Simulator timeframe. b) the Observer timeframe. The labels F, C, and
S refers respectively to flat, ``chess-board-like'', and ``strips-like''
spatiotemporal patterns.
}
\end{figure}

Figures 2(a) and 2(b) show, respectively, the
spatiotemporal pattern in the Simulator and in the Observer timeframes
for  $\epsilon = 0.13$.
The spatial patterns has been displayed each 120 iterations in order
to show the long term structure. Note that the spatiotemporal pattern
in the Simulator timeframe looks very irregular, in the sense that nearest
neighbors tend to have dissimilar states. However, some structures
can be observed. 
\begin{figure*}
 \includegraphics[width=14cm]{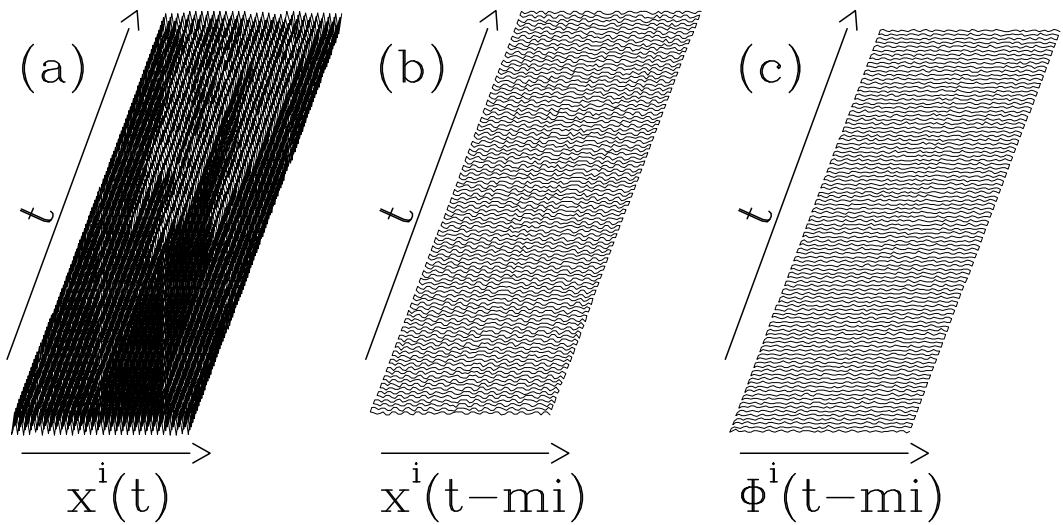}
\caption{ Spatiotemporal pattern at 120 iterations interval in
a) the Simulator timeframe. b) the Observer timeframe. c) spatiotemporal pattern
in the Observer timeframe of the forcing field $\Phi^i(t-m|i-i_c|)$
}
\end{figure*}
\begin{figure*}
 \includegraphics[width=14cm]{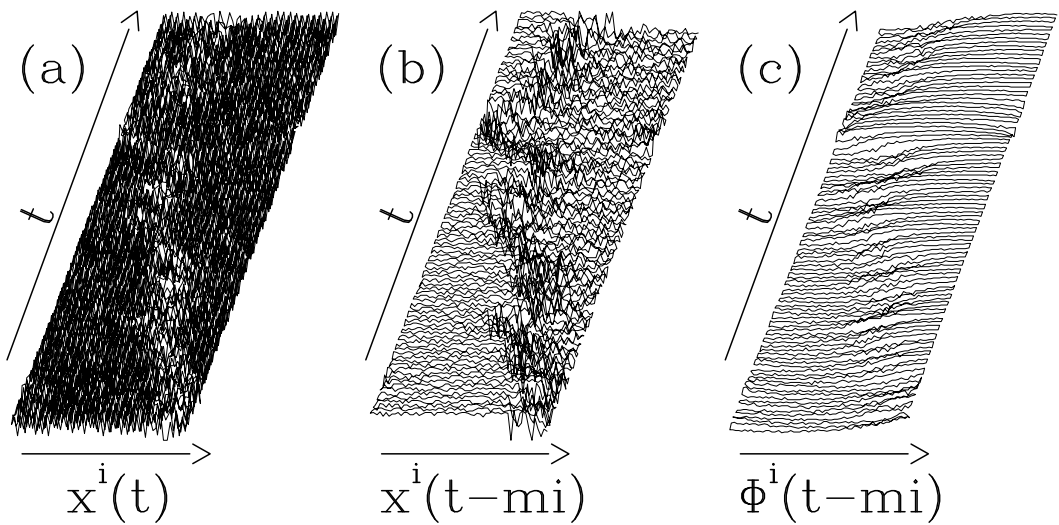}
\caption{ Same as FIG 2 but for $N=59$.
}
\end{figure*}
On the other hand, the spatiotemporal pattern
in the Observer timeframe (i.e. the system as seen by the central 
element $i_c$) 
looks smooth and much more organized.
Figure 2(c) shows the spatiotemporal pattern of the forcing field 
$\Phi^i(t-m|i-i_c|)$ in the Observer timeframe. Note that the pattern in
figure 2(c) is even smoother that the pattern in figure 2(b).
To ilustrate the drastic effect that might have the ratio $N/N_v$ figures
3(a-c) show the patterns in figures 2(a-c) but for the case with $N=59$. 

Figures 4(a) and 4(b) show the return map of the spatial mean value $<x>$
in the Simulator and Observer timeframes, respectively. Due to the
spatial irregularity in the Simulator timeframe, the mean $<x>_{\rm Sim}$
remains about constant and occupies a small area in the return plane.
However, in the Observer timeframe the mean  
$<x>_{\rm Obs}$ displays almost a period 2.
A similar contrast is observed in the return map
of the spatial dispersion $\sigma$.
\begin{figure}
 \includegraphics[width=8cm]{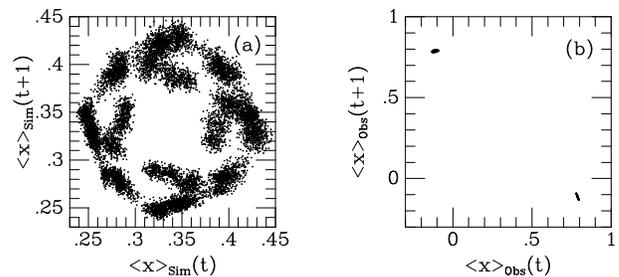}
\caption{ Return of the spatial mean value in
a) the Simulator timeframe. b) the Observer timeframe.
}
\end{figure}

Note that in Eq. (1) the delay is included in the forcing field
$\Phi$. If the value of $x^i(t)$ is systematically 
correlated with the value of $\Phi^i(t)$, then we expect
to have a smooth spatiotemporal pattern in the Observer 
timeframe. Figures 5(a) and 5(b) show, respectively, the 
return map of the forcing $\Phi^1(t)$ of the first element
and the forcing $\Phi^1(t)$ VS the state of the first element.
Note that in figure 5(b) the points remain close to the diagonal.
For cases where the spatiotemporal pattern in the Observer
timeframe looks irregular, the corresponding figure 5(b) would show
an anticorrelation or a poor correlation between $x^i(t)$ 
and $\Phi^i(t)$.
\begin{figure}
 \includegraphics[width=8cm]{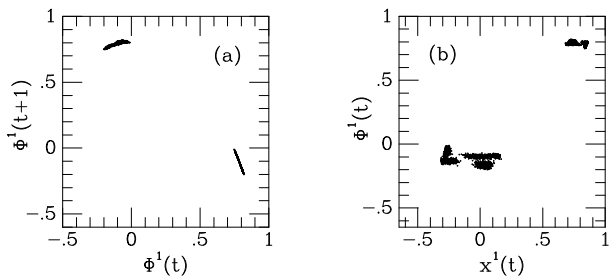}
\caption{ a) Return of the first element forcing $\Phi^1(t)$.
b) Forcing versus state on the first element. 
}
\end{figure}

For the same parameter values in the above example, figures 6(a-d) show 
the return map of the spatial mean value $<x>_{Sim}$ in
the Simulator timeframe for various time delays: a) $m=3$, b) $m=7$, c) $m=11$,
and d) $m=15$. For odd values of $m$, the return map of the spatial mean value 
$<x>_{Obs}$ in the Observer timeframe looks almost the same as the one in figure 4(b),
which correspond to $m=1$. In contrast, the shape of the returns in the 
Simulator timeframe show a clear dependence on the time delay $m$, demonstrating
the convenience of analyzing the collective behavior of the system 
in the Observer timeframe. 
\begin{figure}
 \includegraphics[width=8cm]{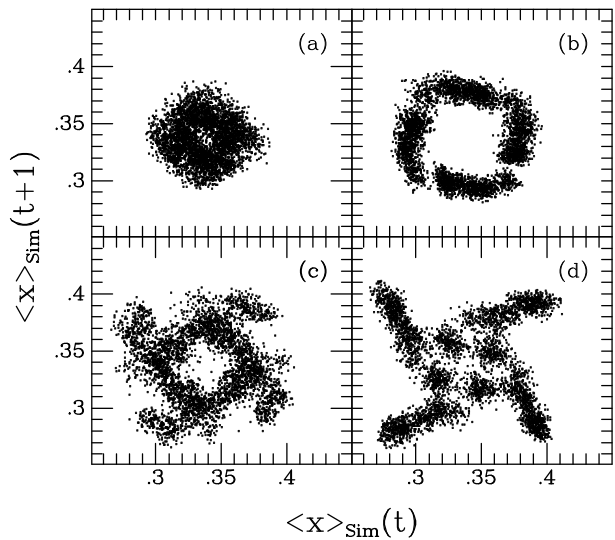}
\caption{ Return of the spatial mean value in
the Simulator timeframe for various time delays. a) $m=3$, b) $m=7$, c) $m=11$,
and d) $m=15$. 
}
\end{figure}

The system seen in the Observer timeframe does not necessarily
show more coherence than in the Simulator one. 
Nevertheless, the coherence observed in the Simulator timeframe is in all cases
the result of the subjacent causality in the spatiotemporal pattern
in the Observer timeframe.

In summary, the presence of delayed interactions in the
CMN defined by Eqs. (1) and (3) induces
a variety of organized behavior. It is out of the 
scope of this short paper to explore and characterize the
model outputs in the parameter space of the system . Here, we just
mention that smooth spatiotemporal patterns in the Observer timeframe
are generally associated with small values of the spatial dispersion and 
the difference $|x^i(t)-\Phi^i(t)|$.
In any case, we emphasize the convenience of looking at the evolution
of networks with delayed interactions in the Observer timeframe since
this is a causal timeframe. Additionally, we have shown that relevant
information can be obtained by
analyzing the dynamics of the local coupling field $\Phi$ since it contains 
the sum of all the delayed interactions.
The use of the Observer timeframe can also be relevant for the analysis
of experimental data and of models of continuous spatiotemporal systems.

\section*{ Acknowledgments}
We thanks David Hollenbach, Mario Cosenza, Jos\'e Albornoz and Juli\'an Suarez
for useful comments that helped to improve the manuscript.
The research of AP is supported by the University of Los Andes 
(CDCHT project C-1285-04-05-A).

\end{document}